\documentclass[twocolumn,showpacs,preprintnumbers]{revtex4-1}
\usepackage{fancyvrb}
\usepackage{multimedia}
\usepackage{subfigure}
\usepackage{amssymb,amsmath}
\def \be{\begin{equation}}
\def \ee{\end{equation}}
\def \bdm{\begin{eqnarray}}
\def \edm{\end{eqnarray}}
\begin{document}
\preprint{Submitted to Physics of Plasmas}
\title{Perpendicular Diffusion of Energetic Particles in Collisionless Plasmas}
\author{A. Shalchi}
\email{andreasm4@yahoo.com}
\affiliation{Department of Physics and Astronomy, University of Manitoba, Winnipeg, Manitoba R3T 2N2, Canada}
\date{\today}
\begin{abstract}
A fundamental problem in plasma and astrophysics is the interaction between energetic particles and magnetized
plasmas. In the current paper we focus on particle diffusion across the guide magnetic field. It is shown that
the perpendicular diffusion coefficient depends only on the parallel diffusion coefficient and the Kubo number.
Therefore, one can find four asymptotic limits depending on the values of these two parameters.
These regimes are the quasilinear limit, the Kadomtsev \& Pogutse limit, the scaling of Rechester
\& Rosenbluth, and the scaling found by Zybin \& Istomin. In the current article we focus on the Rechester
\& Rosenbluth scenario because this was not discovered before in the context of collisionless plasmas.
Examples and applications are discussed as well. We show that an energy independent ratio of perpendicular and
parallel diffusion coefficients can be found and that this ratio can be very small but also close to unity.
This is exactly what one observes in the solar wind.
\end{abstract}
\maketitle
\section{Introduction}
A fundamental problem in physics is the stochastic motion of particles. Originally the stochastic motion
of {\it Brownian particles} was investigated by R. Brown and A. Einstein (see Refs. \cite{brown1827} and
\cite{einstein1905}). Later S. Chandrasekhar presented a more general discussion of stochastic problems in
Physics and Astronomy (see Ref. \cite{chandra43}). In the recent decades the interaction between energetic
particles and a turbulent magnetized plasmas became a topic of great interest (see, e.g., Ref. \cite{Balescu05}).
An interesting example is the motion of alpha particles or accelerated ion beams in a {\it Tokamak} (see, e.g., Ref. \cite{Wesson04}).
In space science and astrophysics, cosmic rays propagate through the plasma of the interstellar space or the
solar wind. In their report to the {\it National Academy of Science}, Turner et al. (see Ref. \cite{Turner2002})
listed the eleven fundamental science questions for the 21st century. One of these questions is the nature of
cosmic ray sources. One ingredient to the solution of this puzzle is the understanding of energetic particle diffusion
because this plays a crucial role in shock acceleration scenarios.

If charged particles move through the plasma, they experience scattering due to the interaction with turbulent
magnetic fields. In addition to such stochastic magnetic fields, in the following represented by the symbol $\delta \vec{B}$,
one also finds a non-vanishing mean (guide) magnetic field $\vec{B}_0 = B_0 \vec{e}_z$. Since the latter field breaks
the symmetry of the physical system, one has to distinguish between diffusion of particles along and across
the mean field. In analytical treatments of the transport, it appears that perpendicular diffusion is more difficult
to understand (see, e.g., Ref. \cite{shal09book}).

Recently the so-called {\it Unified Non-Linear Transport (UNLT) theory} was presented in Ref. \cite{shalapjlett}.
The validity of the latter theory was tested by comparing its results with test-particle simulations. Remarkable
agreement was found for a variety of turbulence models such as the slab/2D model and nosiy reduced magnetohydrodynamic (MHD)
turbulence (see Refs. \cite{tausha2011} and \cite{shalhuss14}). In Ref. \cite{shal2013} it was shown that
UNLT theory agrees also with test-particle simulations performed for models based on Goldreich \& Sridhar
scaling (see Ref. \cite{gold95}).

In the current paper we show that the perpendicular diffusion coefficient depends only on two parameters,
namely the parallel mean free path and the Kubo number. Furthermore, a new formula for the perpendicular
diffusion coefficient is derived from UNLT theory corresponding to the scaling obtained by Rechester \& Rosenbluth
(see Refs. \cite{spatrev08} and \cite{rechrosen78}). After discussing examples, it is argued that the
latter transport regime is important for explaining solar wind observations.
\section{Perpendicular Diffusion}
The UNLT theory provides the following nonlinear integral equation for the perpendicular diffusion coefficient
\be
\kappa_{\perp} = \frac{a^2 v^2}{3 B_0^2} \int d^3 k \; \frac{P_{xx} (\vec{k})}{F(k_{\parallel},k_{\perp}) + (4/3) \kappa_{\perp} k_{\perp}^2 + v/\lambda_{\parallel}}
\label{UNLT}
\ee
with $F(k_{\parallel},k_{\perp}) = (v^2 k_{\parallel}^2)/(3 \kappa_{\perp} k_{\perp}^2)$. Here we have used
the parallel and perpendicular components of the wave vector $k_{\parallel}$ and $k_{\perp}$, the magnetic correlation
tensor $P_{mn} (\vec{k}) = \langle \delta B_m (\vec{k}) \delta B_n^{*} (\vec{k}) \rangle$, the parallel diffusion coefficient of
the particle $\kappa_{\parallel}$, the parallel mean free path $\lambda_{\parallel} = 3 \kappa_{\parallel} /v$, and the
particle speed $v$. We have also used the order one parameter $a^2$ as in Ref. \cite{matt03}. Eq. (\ref{UNLT}) is valid
for axis-symmetric static turbulence, $\delta B_z \ll B_0$, and a constant guide field.

In the following we assume that the combined correlation function $P (\vec{k})=P_{xx} (\vec{k})+P_{yy} (\vec{k})$
has the form 
\be
P \left( \vec{k} \right) = l_{\parallel} l_{\perp}^2 \delta B_x^2 \; f \left( k_{\parallel} l_{\parallel}, k_{\perp} l_{\perp} \right)
\label{turbulence}
\ee
with the dimensionless function $f(x, y)$ which depends only on $x=k_{\parallel} l_{\parallel}$
and $y=k_{\perp} l_{\perp}$. Furthermore, this function decays with increasing parallel and perpendicular wavenumbers. Therefore,
$l_{\parallel}$ and $l_{\perp}$ are characteristic length scales for the decorrelation of the turbulence. The form (\ref{turbulence})
was chosen so that $\int d^3 k \; P (\vec{k})=\delta B^2 = \delta B_x^2 + \delta B_y^2$. Therefore, the function $f$
satisfies the normalization condition $1 = 2 \pi \int_{0}^{\infty} d y \int_{0}^{\infty} d x \; y f(x,y)$.  
To proceed we use the diffusion ratio
$D := (\kappa_{\perp} l_{\parallel}^2)/(\kappa_{\parallel} l_{\perp}^2) = (\lambda_{\perp} l_{\parallel}^2)/(\lambda_{\parallel} l_{\perp}^2)$ and the Kubo number
\be
K = \frac{l_{\parallel}}{l_{\perp}} \frac{\delta B_x}{B_0}.
\label{Kubo}
\ee
Therewith, Eq. (\ref{UNLT}) can be written as
\bdm
D & = & 2 \pi a^2 K^2 \int_{0}^{\infty} d y \; \int_{0}^{\infty} d x \; y f(x,y) \nonumber\\
& \times & \frac{D}{x^2/y^2 + D + \left( 2 D \lambda_{\parallel} \right)^2/\left( 3 l_{\parallel} \right)^2 y^2}.
\label{unltkubo1}
\edm
Obviously, there are only two parameters controlling the diffusion ratio $D$, namely the parallel mean free path normalized to the
parallel scale $\lambda_{\parallel}/l_{\parallel}$ and the Kubo number $K$. We conclude that there are four different asymptotic limits
and therewith four different transport regimes. Those are discussed in the following. All limits are compared with each other in
Table \ref{casetable2}.
\subsection{The quasilinear regime}
First we consider the limit $\lambda_{\parallel}/l_{\parallel} \rightarrow \infty$ corresponding to the case that pitch-angle scattering
and therewith parallel diffusion are suppressed. In this case it is convenient to go back to Eq. (\ref{UNLT}) and set $v/\lambda_{\parallel} = 0$.
In this limit Eq. (\ref{UNLT}) has the solution $\kappa_{\perp} = v \kappa_{FL}/2$ where the field line diffusion coefficient $\kappa_{FL}$
is given by 
\be
\kappa_{FL}
= \frac{1}{\tilde{B}_0^2} \int d^3 k \; P_{xx} \left( \vec{k} \right) \frac{\kappa_{FL} k_{\perp}^2}{k_{\parallel}^2 + \left( \kappa_{FL} k_{\perp}^2 \right)^2}
\label{bill1}
\ee
in agreement with the theory developed by Matthaeus et al. (see Ref. \cite{matt95}). Here we have introduced the effective guide field
$\tilde{B}_0 = B_0 / a$.
By using again the form (\ref{turbulence}) and by introducing the dimensionless field line
diffusion coefficient $\tilde{D} = \kappa_{FL} l_{\parallel} / l_{\perp}^2$, Eq. (\ref{bill1}) becomes
\be
\tilde{D} = 2 \pi K^2 \int_{0}^{\infty} d y \; \int_{0}^{\infty} d x \; y f(x,y) \frac{\tilde{D} y^2}{x^2 + \left( \tilde{D} y^2 \right)^2}.
\label{flkubo}
\ee
We find that $\tilde{D}$ depends only on one single parameter and that is the Kubo number $K$.
One can obtain two different asymptotic solutions of the latter integral equation by considering small
and large Kubo numbers.
In the limit $K \rightarrow 0$ we expect $\tilde{D} \rightarrow 0$. By using the relation $\lim_{\xi \rightarrow 0} \xi/(x^2 + \xi^2)=\pi \delta (x)$, with the Dirac delta $\delta (x)$, Eq. (\ref{flkubo}) becomes
\be
\tilde{D} = 2 \pi^2 K^2 \int_{0}^{\infty} d y \; y f (x=0,y)
\ee
and the parameter $\tilde{D}$, and therewith the field line diffusion coefficient $\kappa_{FL}$, scales with $K^2$
corresponding to the well-known quasilinear scaling (see, e.g., Ref. \cite{jokpar69}).
\subsection{The Kadomtsev \& Pogutse scaling}
In the limit of large Kubo numbers $K \rightarrow \infty$ we expect $\tilde{D} \rightarrow \infty$. In this case Eq. (\ref{flkubo}) becomes
\be
\tilde{D}^2 = 2 \pi K^2 \int_{0}^{\infty} d y \; \int_{0}^{\infty} d x \; y^{-1} f (x,y)
\ee
and we find the scaling $\tilde{D} \propto K$. The latter limit is sometimes called the nonlinear regime or the Bohm limit of field line
diffusion and was originally obtained by Kadomtsev \& Pogutse (see Ref. \cite{KadPog78}). 
\subsection{The Zybin \& Istomin scaling}
In the limit $\lambda_{\parallel}/l_{\parallel} \rightarrow 0$, Eq. (\ref{unltkubo1}) becomes
\be
D = 2 \pi a^2 K^2 \int_{0}^{\infty} d y \; \int_{0}^{\infty} d x \; y f(x,y) \frac{D}{x^2/y^2 + D}.
\label{strongpas}
\ee
For large Kubo numbers we expect that the diffusion parameter $D$ becomes large as well and Eq. (\ref{strongpas})
simplifies to
\be
D = 2 \pi a^2 K^2 \int_{0}^{\infty} d y \; \int_{0}^{\infty} d x \; y f(x,y).
\label{unltkubo2}
\ee
Obviously $D \propto K^2$. Eq. (\ref{unltkubo2}) can be rewritten as $\kappa_{\perp}=a^2 \kappa_{\parallel} \delta B_x^2 / B_0^2$.
The latter result is also known in diffusion theory (see Refs. \cite{ZybIsto85} and \cite{shal09book}) and we refer to it as the
{\it Zybin \& Istomin scaling}.
\subsection{The collisionless Rechester \& Rosenbluth scaling}
For small Kubo numbers we use the relations $\lim_{\xi \rightarrow 0} \xi/(x^2 + \xi^2)=\pi \delta (x)$ and
$\delta \left( \alpha z \right) = \delta (z) / |\alpha|$ to derive from Eq. (\ref{strongpas})
\be
\sqrt{D} = 2 \pi^2 a^2 K^2 \int_{0}^{\infty} d y \; y^2 f(x=0,y).
\label{newmode}
\ee
Assuming that the $y$-integral is convergent, we find the scaling $D \propto K^4$. To proceed we define the
dimensionless parameter $\gamma^2 := 2 \pi^2 \int_{0}^{\infty} d y \; y^2 f(x=0,y)$. Furthermore, we replace
the parameters $D$ and $K$ by standard quantities to obtain the scaling
\be
\frac{\kappa_{\perp}}{\kappa_{\parallel}} = \gamma^2 a^4 \frac{l_{\parallel}^2}{l_{\perp}^2} \frac{\delta B_x^4}{B_0^4}.
\label{unltrandr}
\ee
The latter limit was derived from the UNLT theory. In order to understand this result, we compare our finding with the
formulas discussed in Ref. \cite{spatrev08}. According to the latter reference, Rechester \& Rosenbluth proposed
$\kappa_{\perp} \propto (l_{\parallel} \kappa_{\parallel} \delta B_x^2)/(L_K B_0^2)$ with the Kolmogorov length $L_K$.
If the latter parameter is replaced by $L_K = (\sqrt{2} l_{\perp}^2 B_0^2)/(4 \sqrt{\pi} l_{\parallel} \delta B_x^2)$
(see, e.g., Ref. \cite{neuspa06}), one obtains
\be
\frac{\kappa_{\perp}}{\kappa_{\parallel}} = 2 \sqrt{2 \pi} \frac{l_{\parallel}^2}{l_{\perp}^2} \frac{\delta B_x^4}{B_0^4}
\label{spat08}
\ee
which agrees with Eq. (\ref{unltrandr}).
By replacing $D$ and $K$ by standard parameters, Eq. (\ref{newmode}) can be written as
\be
\sqrt{\frac{\kappa_{\perp}}{\kappa_{\parallel}}}
= \pi \frac{a^2}{B_0^2} \int d^3 k \; P_{xx} \left( \vec{k} \right) k_{\perp} \delta (k_{\parallel}).
\label{standardform}
\ee
The latter formula can easily be combined with any turbulence model as long as the occurring integrals are convergent.
We refer to this limit as the {\it collisionless Rechester \& Rosenbluth (CLRR) scaling}.
\section{Illustrative Examples}
A simple model for turbulence is the {\it slab model} where all wave vectors are assumed to be parallel wrt the mean field.
It is well-known that perpendicular transport is subdiffusive in this case (see, e.g., Ref. \cite{qin2002}). Of course this
subdiffusive behavior does only occur if the parallel mean free path is finite. In the asymptotic limit $\lambda_{\parallel} \rightarrow \infty$,
the usual quasilinear scaling can be obtained as discussed above. For a finite parallel mean free path, we find a transport
process which is usually called {\it compound diffusion} (see Ref. \cite{webb06} for details).
In the following it is shown that diffusion is restored if the slab model is broadened. The broadening of a turbulence model with
reduced dimensionality was discussed before in Ref. \cite{ruffmatt2013}. Physically this corresponds to the case
where wave vectors are mainly oriented parallel wrt the mean field but weak fluctuations are taken into account.
We refer to this model as the {\it noisy slab model} and we define it via
\be
P \left( \vec{k} \right) = \frac{2 l_{\perp}}{k_{\perp}} g (k_{\parallel}) \Theta \left( 1 - k_{\perp} l_{\perp} \right)
\label{noisyslab}
\ee
where we have used the {\it Heaviside step function} $\Theta (x)$. For the spectrum of the slab modes we employ
$g (k_{\parallel}) = C(s)/(2 \pi) \delta B^2 l_{\parallel} \left[ 1 + (k_{\parallel} l_{\parallel})^2 \right]^{-s/2}$
(see, e.g., Ref. \cite{bieber94}). The normalization function $C(s) = \Gamma (s/2) / [2 \sqrt{\pi} \Gamma ((s-1)/2)]$ depends
on the inertial range spectral index $s$ and Gamma functions. By combining the noisy slab model with Eq. (\ref{standardform}),
one obtains
\be
\frac{\kappa_{\perp}}{\kappa_{\parallel}} = \left[ \frac{\pi}{2} C(s) a^2 \frac{l_{\parallel}}{l_{\perp}} \frac{\delta B^2}{B_0^2} \right]^2.
\label{finalnoisy}
\ee
We can recover the case of pure slab turbulence by setting $l_{\perp} = \infty$ and, thus, we find $\kappa_{\perp}=0$ corresponding
to subdiffusion.
To estimate a number for the ratio $\kappa_{\perp} / \kappa_{\parallel}$, we use $\delta B^2 / B_0^2 \approx 0.5$,
$l_{\parallel}/l_{\perp} \approx 0.75$, and $a^2 \approx 1$. For $s=5/3$ (see Ref. \cite{kol41}) we find $C(s=5/3) \approx 0.12$
and, thus, Eq. (\ref{finalnoisy}) provides $\kappa_{\perp}/\kappa_{\parallel} \approx 0.005$.

As a second example we employ the Gaussian decorrelation model used in Ref. \cite{neuspa06} which is given by
\be
P \left( \vec{k} \right)=\frac{l_{\parallel} l_{\perp}^{4} \delta B_x^2}{(2 \pi)^{3/2}} k_{\perp}^2
e^{-\frac{1}{2} (l_{\parallel} k_{\parallel})^2 -\frac{1}{2} (l_{\perp} k_{\perp})^2}.
\label{neuerspek}
\ee
A problem of this model is the rapid decorrelation if the wave numbers are larger than the inverse scales.
This behavior is not realistic in astrophysical scenarios such as the solar wind. In reality one expects
a power law behavior with $k^{-5/3}$ (see Ref. \cite{kol41}). Such small scales, usually referred to as
inertial range, and even smaller scales, are essential if it comes to pitch-angle scattering and
therewith parallel diffusion (see, e.g., Refs. \cite{bieber94} and \cite{shal09book}). In the
context of fusion plasmas, however, the form (\ref{neuerspek}) is commonly used.
Combining formula (\ref{standardform}) with the spectrum (\ref{neuerspek}), one finds
\be
\frac{\kappa_{\perp}}{\kappa_{\parallel}}
= \left( \frac{3 \pi}{4} \right)^2 a^4 \frac{l_{\parallel}^2}{l_{\perp}^2} \frac{\delta B_x^4}{B_0^4}.
\ee
The latter result is in agreement with the form (\ref{unltrandr}) derived above. Furthermore, we can compare our finding with
Eq. (\ref{spat08}). We find agreement if we set $a^2 = \pi^{-3/4} 2^{11/4} / 3 \approx 0.95$ confirming the assumption that
$a^2$ is indeed an order one constant.

\begin{table*}
\caption{Transport regimes described by UNLT theory for extreme values of the parallel mean free path wrt the parallel scale
$\lambda_{\parallel}/l_{\parallel}$ and the Kubo number $K$. For each case we have listed the formula for the perpendicular
diffusion coefficient $\kappa_{\perp}$ and the corresponding scaling behavior. All used parameters are explained in the main
part of the text.}
\begin{center}
\renewcommand{\arraystretch}{1.5}
{\large
\begin{tabular}{|l|l|l|l|}
\hline
Transport Regime				& Limit											& Perpendicular Diffusion Coefficient			& Scaling												\\
\hline

Quasilinear Diffusion			& $\lambda_{\parallel}/l_{\parallel} \rightarrow \infty$, $K \rightarrow 0$		& $\kappa_{\perp} = \frac{\pi}{2} \frac{v a^2}{B_0^2} \int d^3 k \; P_{xx} (\vec{k}) \delta (k_{\parallel})$& $\lambda_{\perp} \propto l_{\parallel} \frac{\delta B_x^2}{B_0^2}$	\\

Kadomtsev \& Pogutse		& $\lambda_{\parallel}/l_{\parallel} \rightarrow \infty$, $K \rightarrow \infty$	& $\kappa_{\perp} = \frac{v a}{2 B_0} \left[ \int d^3 k \; P_{xx} (\vec{k}) k_{\perp}^{-2} \right]^{1/2}$	& $\lambda_{\perp} \propto l_{\perp} \frac{\delta B_x}{B_0}$			\\

Rechester \& Rosenbluth		& $\lambda_{\parallel}/l_{\parallel} \rightarrow 0$, $K \rightarrow 0$			& $\kappa_{\perp} = \frac{\pi^2 \kappa_{\parallel} a^4}{B_0^4} \left[ \int d^3 k \; P_{xx} (\vec{k}) k_{\perp} \delta (k_{\parallel}) \right]^{2}$		& $\lambda_{\perp} \propto \lambda_{\parallel} \frac{l_{\parallel}^2}{l_{\perp}^2} \frac{\delta B_x^4}{B_0^4}$		\\

Zybin \& Istomin		& $\lambda_{\parallel}/l_{\parallel} \rightarrow 0$, $K \rightarrow \infty$		& $\kappa_{\perp} = \frac{a^2 \kappa_{\parallel}}{B_0^2} \int d^3 k \; P_{xx} (\vec{k}) \equiv a^2 \kappa_{\parallel} \frac{\delta B_x^2}{B_0^2}$		& $\lambda_{\perp} \propto \lambda_{\parallel} \frac{\delta B_x^2}{B_0^2}$	\\
\hline
Compound Diffusion			& $\lambda_{\parallel}/l_{\parallel}$ finite, $K=0$								& $\frac{1}{2} \frac{d}{d t} \left( \left< \Delta x \right> \right)^2 = \kappa_{FL} \sqrt{\kappa_{\parallel}/(\pi t)}$	&	Subdiffusive		\\
\hline
\end{tabular}
}
\end{center}
\label{casetable2}
\end{table*}
\section{Summary and Conclusion}
We have shown that the perpendicular diffusion coefficient of energetic particles interacting with turbulent magnetic fields
depends only on two parameters, namely the ratio $\lambda_{\parallel}/l_{\parallel}$ and the Kubo
number $K$. By considering asymptotic limits, we can identify four transport regimes, namely the
{\it quasilinear regime}, the {\it Kadomtsev \& Pogutse limit}, the {\it Zybin \& Istomin scaling}, and the
{\it collisionless Rechester \& Rosenbluth (CLRR) scaling}. All cases are summarized in Table \ref{casetable2}.
There we also list the corresponding formulas for the perpendicular diffusion coefficient and the scaling behavior.

In our work we focus on the {\it CLRR scaling} which was derived the first time from UNLT theory. This transport
regime is obtained for short parallel mean free paths and small Kubo numbers. In this case the ratio
$\kappa_{\perp} / \kappa_{\parallel}$ is constant and very small. Requiring convergence of the
corresponding wavenumber integral, the perpendicular diffusion coefficient is in this case given by
Eq. (\ref{standardform}). We like to point out that a collisionless adaption of Rechester \& Rosenbluth was
already presented in Ref. \cite{Chan98}. The letter result, however, disagrees with the formula derived in the
current article and simulations of test-particle transport (see, e.g., \cite{matt03}).

A characteristic aspect of the quasilinear and the {\it Kadomtsev \& Pogutse} regimes is the fact that the
perpendicular diffusion coefficient does not depend on the magnetic rigidity whereas in the other two cases,
the ratio $\kappa_{\perp}/\kappa_{\parallel}$ is energy independent and the Kubo number controls
the magnitude of the ratio $\kappa_{\perp}/\kappa_{\parallel}$. For large Kubo numbers $\kappa_{\perp}/\kappa_{\parallel}$ is
close to unity as long as $\delta B \approx B_0$. For small Kubo numbers, however, the ratio $\kappa_{\perp}/\kappa_{\parallel}$
can be very small. A constant value of $\kappa_{\perp}/\kappa_{\parallel}$ was discussed before in the context of solar wind
observations (see Ref. \cite{bieber04} for more details). However, from observations, it is not entirely clear what the
value of this ratio really is. It is often assumed that it is very small (e.g., $\kappa_{\perp}/\kappa_{\parallel} \approx 0.01$
or even smaller) but there are also reports that $\kappa_{\perp}/\kappa_{\parallel} \approx 1$ (see Refs. \cite{dwyer97}
and \cite{zhang03}). In the current paper an explanation for this behavior is provided.
\begin{acknowledgments}
{\it Andreas Shalchi acknowledges support by the Natural Sciences and Engineering Research Council (NSERC) of Canada.}
\end{acknowledgments}
{}

\end{document}